\documentclass[aps,prl,preprint,showpacs,superscriptaddress,groupedaddress]{revtex4}
\usepackage{amsmath}
\usepackage{tipa}
\usepackage{bbm}
\usepackage{txfonts}
\usepackage{graphicx}
\usepackage{dcolumn}
\usepackage{bm}
\usepackage{amssymb}
\usepackage{latexsym}
\usepackage{color}
\usepackage[marginal]{footmisc}

\begin{document}

\title{Nanoscale angular lattice formed by light with high orbital angular momentum}

\author{Zheng Li \footnote{These authors have contributed equally to this work.} \footnote{zheng.li@desy.de} } 
\affiliation{Max Planck Institute for the Structure and Dynamics of Matter, 22761 Hamburg, Germany}
\affiliation{Center for Free-Electron Laser Science, Deutsches Elektronen-Synchrotron DESY, Notkestrasse 85, 22607 Hamburg, Germany}

\author{Xiang Gao\textsuperscript{$\ast$ }}
\author{Lu-Shuai Cao \footnote{lushuai\_cao@hust.edu.cn} }
\affiliation{MOE Key Laboratory of Fundamental Physical Quantities Measurement ${\rm{\&}}$ Hubei Key Laboratory of Gravitation and Quantum Physics, PGMF and School of Physics, Huazhong University of Science and Technology, Wuhan 430074, P. R. China}

\date{\today}

\begin{abstract}

Standing waves generated by the interference of Laguerre-Gaussian (LG) beams can be used for dipole trap. We propose a scheme to create a nanometer (nm) scale ring lattice based on the interference of two high order LG beams without decrease the wavelength. Both of the two LG beams have a monocyclic intensity distribution, and they have the same orbital angular momentum (OAM) quantum number on the order of $10^6$. We are able to theoretically demonstrate a dipole potential along angular direction with the period of $\sim$1 nm , given the waist of the Gaussian beams to be $\sim$0.8 $\mu$m. The atoms in this lattice can be trapped along the radial and azimuthal direction in the potential wells of with $\sim$100 nm and $\sim$0.85 nm distance. The proposed method opens up a convenient pathway towards sub-wavelength atom traps that could directly lead to overlap of wave function of atoms in adjacent wells and the formation of molecular bonds.

\end{abstract}


\maketitle
\noindent
\textbf{I. Introduction}

The angular momentum of light can be divided into the spin part and orbital part \cite{Bekshaev11}, while the spin angular momentum (SAM) that is related to the polarization of the light, and the orbital part is related to the orbital angular momentum (OAM)of light beams. Since firstly found by L. Allen in 1992, that laser with Laguerre-Gaussian (LG) amplitude distribution have a well-defined OAM \cite{Allen92}, light and particle beams with OAM have drawn attention of researchers in various fields \cite{Lloyd17,Devlin17}, such as optical tweezer based on transferring OAM from light to matter particles \cite{He95,MacDonald02}, atom guiding and trapping \cite{Wright00}, and exotic excitation of atoms \cite{Scholz14,Rodrigues16,Afanasev14}.

The angular lattice, which is formed by standing waves originated from the interference of two LG beams with nonzero OAM, can also be used into atom trapping \cite{Franke07,Huang16,Rhodes06}. The period of interference mode in angular direction is proportional to {$\pi/L$}, where $L$ is the OAM quantum number of the LG beam, and is not limited by the wavelength of LG beams. This can be used as novel way of creating the small period lattice without demanding short wavelength of the laser.

On the other hand, it is of crucial importance in ultracold atom physics to realize optical lattice beyond the diffraction limit. In 2013, Cirac et al. proposed a magnetic lattice for ultracold atoms on the scale of few tens of nanometers \cite{Nascimbene15,Wang18,Romero13}. However, the smallest period of the proposed sub-wavelength optical lattices have not yet reached sub-10 nm scale, in such length scale, the wave functions of atoms in adjacent traps could overlap. This can lead to exciting new physics, such as direct observation of molecular bond formation in such lattices and creation of novel molecular structures \cite{LRLiu13}.

There are various methods to generate OAM carrying light beams \cite{McGloin05,Guti谷rrez00,Bandres04}, e.g. using spiral phase plate (SPP) and spiral phase mirrors (SPM) as mode converter \cite{Beijersbergen93,Heckenberg92,Turnbull96,Arlt98,Fickler16} and holographic method \cite{Courtial97}. The latter method avoids using thick lenses and can produce LG beam with OAM exceeding $10^4\hbar$. According to the quantum theory, the orbital angular momentum can, in principle, be arbitrarily large \cite{Fickler16}.

In this work, we theoretically present a novel scheme to create sub-nm scale optical lattices based on the high order LG beams. We demonstrate that an angular optical dipole lattice can be obtained from IR laser with $\sim$1 nm period, which is close to the scale of lattices in realistic solid state materials. The angular lattice formed by such LG beams can be more benign to atoms, while the lattice of x-ray laser with the similar wavelength must destroy the structure of the atoms and molecules within 100 fs through photoionization and subsequent Coulomb explesion \cite{Barty12}.

This work is organized as follows: In Section II, we introduce the related properties of the LG beams and calculate the optical dipole potential (ODP) generated by the interference of LG beams. In Section III, we solve the ground state of a single atom in this ODT. Finally, in Section IV, we present the wave function of single $^{87}$Rb atom in this angular lattice formed by two LG beams with wavelength of 780 nm.

\noindent
\textbf{II.	The Laguerre Gaussian beams and the optical dipole potential}

The circularly symmetric LG beam is denoted by ${\rm{LG}}_p^L$  where $L$ is the quantum number of azimuthal mode, and the beams with $L \ne 0$ give rise to a well-defined OAM of $L\hbar$ per photon \cite{Allen92}. $p$ is the quantum number of radial mode with $p+1$ nodes.

The normalized field amplitude distribution of an ${\rm{LG}}_p^L$ mode laser beam can be written as \cite{Clifford98}
\begin{equation}
\begin{aligned}
u_{L,p}=\sqrt {\frac{{2p!}}{{\pi \left( {p + \left| L \right|!} \right)}}} \frac{{\sqrt {{P_0}} }}{{w\left( z \right)}}{\left( {\frac{{r\sqrt 2 }}{{w\left( z \right)}}} \right)^{\left| L \right|}}\exp \left( { - \frac{{{r^2}}}{{{w^2}\left( z \right)}}} \right)\left\{ {L_p^{\left| L \right|}\left( {\frac{{2{r^2}}}{{{w^2}\left( z \right)}}} \right)} \right\}\\
\times\exp \left( { - \frac{{ik{r^2}z}}{{2\left( {{z^2} + z_R^2} \right)}}} \right)\exp \left( { - iL\phi } \right)\exp \left[ {i\left( {2p + L + 1} \right){{\tan }^{ - 1}}\left( {\frac{z}{{{z_R}}}} \right)} \right]\,\label{Eq 1}
\end{aligned}
\end{equation}

\noindent
where $z$ is the longitudinal distance from the beam waist (see Fig. 1). $P_0$ is the power of laser beams, $w_z$is the radius at which the Gaussian beam intensity falls to $1/e$ of its axis value while $w_0$ is the beam waist at $z=0$, $L_p^{\left| L \right|}(x)$  is the associated Laguerre polynomial, $\phi$ is the azimuthal angle, ${z_R} = {{\pi w_0^2} \mathord{\left/{\vphantom {{\pi w_0^2} \lambda }} \right.\kern-\nulldelimiterspace} \lambda }$ is the Rayleigh range for the laser with wavelength $\lambda$, and $\tan^{ - 1}\left( {{z \mathord{\left/{\vphantom {z {{z_R}}}} \right.\kern-\nulldelimiterspace} {{z_R}}}} \right)$ is the Gouy phase. We focus on the case of $p=0$, where the associated Laguerre polynomial $L_0^{\left| L \right|}(x) = {\rm{1}}$ and only a single ring is present in the beam intensity profile with the maximum intensity at the toroid radius \cite{Arlt00}
\begin{equation}
r_L = w(z)\sqrt {L/2},\label{Eq 2}
\end{equation}

This maximum intensity is reached at the waist ($z=0$) and is given by
\begin{equation}
{I_{l,\max }} = \frac{2}{{\pi L!}}\frac{{{P_0}}}{{w_0^2}}{L^L}\exp \left( { - L} \right) = \frac{{{P_0}}}{{\pi r_L^2}}\frac{{{L^{L + 1}}\exp \left( { - L} \right)}}{{L!}}\,\label{Eq 3}
\end{equation}

Using Stirling＊s formula, we can further approximate Eq.(3) to \cite{Wright00,Rhodes06}:
\begin{equation}
{I_{L,\max }} = \frac{{{P_0}}}{{\pi r_L^2}}\sqrt {\frac{L}{{2\pi }}} ,\label{Eq 4}
\end{equation}

As the result of fixed toroid radius $r_L$, when the maximum value of intensity increases with the azimuthal index $L$, the radial intensity profile of the beam must become sharper (see Fig. 1 (b), showing the normalized intensity $I\left( r \right)/{I_{1,\max }}$ versus normalized radius ${r \mathord{\left/ {\vphantom {r {{r_L}}}} \right.\kern-\nulldelimiterspace} {{r_L}}}$ for different azimuthal index with the same laser power).

For a circularly polarized beam propagating in the longitudinal z-direction, the linear momentum density ${\varepsilon _0}\vec E \times \vec B$ could be separated into $\vec p_r, \vec p_{\phi}$, and $\vec p_z$ \cite{Allen00}:
\begin{equation}
\begin{aligned}
{p_r} = {\varepsilon _0}\frac{{\omega krz}}{{\left( {z_R^2 + {z^2}} \right)}}{\left| u \right|^2}{{\hat e}_r}\\
{p_\phi } = {\varepsilon _0}\left[ {\frac{{\omega L}}{r}{{\left| u \right|}^2} - \frac{1}{2}\omega \sigma \frac{{\partial {{\left| u \right|}^2}}}{{\partial r}}} \right]{{\hat e}_\phi }\\
{p_z} = {\varepsilon _0}\omega k{\left| u \right|^2}{{\hat e}_z},\label{Eq 5}
\end{aligned}
\end{equation}

\noindent
where $\vec p_r, \vec p_{\phi}$ ,  and $\vec p_z$ \cite{Allen00} are the linear momentum densities in radial, angular and longitudinal directions, respectively. In particular, the first and the second terms of $\vec p_{\phi}$ relate to the orbital angular momentum (OAM) $L\hbar$ and SAM $\sigma\hbar$ of a single photon \cite{Allen92}.

\begin{figure}[tbp]
\includegraphics[width=1\textwidth]{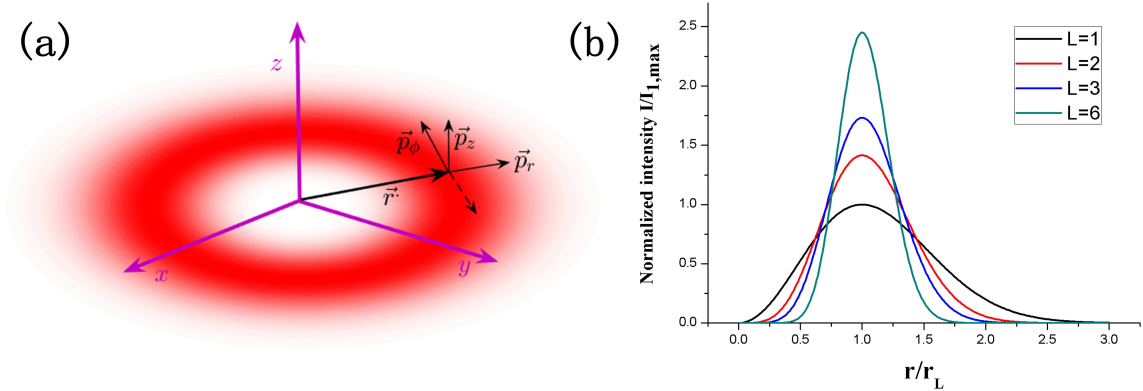}

\caption{\label{fig 1}(color online)  (a) The linear momentum of an LG beams propagating in z-direction with azimuthal index $L>0$. $\vec p_r$ is momentum parallel to the radial vector $\vec r$ and  $\vec p_z$ is the same as wave vector $\vec k$  of the laser while $\vec p_{\phi}$  rotates anti-clockwise with respect to the direction of laser propagation (the solid line), and for the condition that $L<0$ , it is clockwise (the dashed line). (b) Normalized laser intensity $I\left( r \right)/{I_{1,\max }}$  as a function of  $r/r_L$ for $L =1, 2, 3, 6$ respectively with the same toroid radius $r_L$  and the laser power $P_0$.}
\end{figure}

In Fig. 1(a) we show the linear momentum density of a linearly polarized LG beam propagating in the z-direction. For linear polarized beams, we have $\sigma=0$, and counter-rotating momenta $\vec p_{\phi}$ for beams carrying OAM $L$ of opposite signs.

Assuming the laser is linearly polarized, two LG beams propagating in the opposite direction with the same azimuthal index $L$ can interfere, and form standing waves in the angular direction \cite{MacDonald02}. We write the intensity of the interfering beams in the plane of $z=0$ as
\begin{equation}
{I_{{\rm{Interfere}}}}\left( {r,\phi } \right) = 4\sqrt L {I_{1,\max }}{\left( {{r \mathord{\left/
 {\vphantom {r {{r_L}}}} \right.
 \kern-\nulldelimiterspace} {{r_L}}}} \right)^{2L}}{e^{ - L\left( {{r^2}/r_L^2 - 1} \right)}}{\cos ^2}\left( {L\phi } \right),\label{Eq 6}
\end{equation}

\noindent
where ${I_{1,\max }} = {{{P_0}} \mathord{\left/{\vphantom {{{P_0}} {\pi w_0^2\sqrt {2\pi } }}} \right.\kern-\nulldelimiterspace} {\pi w_0^2\sqrt {2\pi } }}$ is the maximum intensity for $L=1$. As one can see from Eq.(7), the interference of ${\rm{LG}}_0^L$  and ${\rm{LG}}_0^{-L}$  beams create an angular lattice with $2L$  nodes in the azimuthal direction.

As one can see from Eq.(6), the intensity distribution of the interference mode depends on two independent variables $r$  and $\phi$. In other words, $I_{\text{Inteference}}(r,\phi)$  can be rewrite as $I_{\text{Inteference}}(r)\times I_{\text{Inteference}}(\phi)$ which is the separation of variables. The line shape of $I_{\text{Inteference}}(\phi)$ is actually $4\cos^2(L\phi)$, while $I_{\text{Interference}}(r)$ has a shape shown in Fig. 1(b),it varies in the form of ${\left( {{r \mathord{\left/{\vphantom {r {{r_L}}}} \right. \kern-\nulldelimiterspace} {{r_L}}}} \right)^{2L}}{e^{ - L\left( {{r^2}/r_L^2 - 1} \right)}}$.

Since the period of the interference mode ${\rm{LG}}_0^{L}+{\rm{LG}}_0^{-L}$ at the beam waist scales as ${s_\phi } = 2{{\pi {r_L}} \mathord{\left/{\vphantom {{\pi {r_L}} {2L}}} \right.\kern-\nulldelimiterspace} {2L}}$, the distance of interval between adjacent traps is not limited by the laser wavelength, and can reach sub-nm scale by already obtainable laser technology.

The atoms in laser fields experience dipole potential resulted from the interaction of induced atomic dipole and light field \cite{Riis,Grimm00}. Though for a laser field with spatial modulation at nm scale, the non-dipole effect can emerge due to inhomogeneity of field intensity in the space in which the atomic and molecular wave functions are distributed, we take here the dipole potential as the first order approximation.

Considering $D_1$ $(5^2S_{1/2}\to 5^2P_{1/2})$ and $D_2$ $(5^2S_{1/2}\to 5^2P_{3/2})$ transitions of the $^{87}$Rb atom in linearly polarized laser field, the ODP can be expressed as \cite{Grimm00}
\begin{equation}
{V_{{\rm{dip}}}}\left( r \right) = \frac{{\pi {c^2}\Gamma }}{{2\omega _0^3}}\left( {\frac{1}{{{\Delta _1}}} + \frac{2}{{{\Delta _2}}}} \right)I\left( r \right),\label{Eq 7}
\end{equation}

\noindent
where  $\Delta_i$ is the detuning of $D_i$  transition, $\Gamma$  is the natural line width of the optical transition and $\omega$ is the central frequency of the laser. The ODP is extremely analogous to the intensity distribution derived from the interference of two LG beams. In this above equation, all of the parameters other than intensity distribution are only related to the atomic properties. For a given atom, the magnitude and shape of the ODT are entirely determined by the intensity distribution of the interference mode.

\noindent
\textbf{III. Bound state in the angular lattice}

For the proposed optical lattice with $\sim$1 nm period, we calculate the bound state of $^{87}$Rb atom inside, the Hamiltonian of which reads
\begin{equation}
H = {T_{r,\phi }} + {V_{{\rm{dip}}}}\left( {r,\phi } \right),\label{Eq 8}
\end{equation}
where the first term refers to the kinetic energy of the atom in polar coordinates ${T_{r,\phi }}{\rm{ = }} - \frac{{{\hbar ^2}}}{{2m}}\left( {\frac{{{\partial ^2}}}{{\partial r_{}^2}} + \frac{1}{r}\frac{\partial }{{{\partial _r}}} + \frac{1}{{{r^2}}}\frac{{{\partial ^2}}}{{\partial \phi _{}^2}}} \right)$, and $V_\text{{dip}}(r,\phi)$  is the dipole potential of the atom in the plane of  $z=0$ i.e. the position of Gaussian beam waist.

We use the perturbation theory to solve the ground state of the system. Firstly, the optical dipole potential in radius direction would be sharper and tighter as $L$  grows. The atom trapped in a vicinity near $r_L$ . We can write the kinetic term approximately as $ T_{r,\phi}\approx - \frac{{{\hbar ^2}}}{{2m}}\left( {\frac{{{\partial ^2}}}{{\partial r_{}^2}} + \frac{1}{{{r_L}}}\frac{\partial }{{{\partial _r}}} + \frac{1}{{r_L^2}}\frac{{{\partial ^2}}}{{\partial \phi _{}^2}}} \right)$ and further as $T_{r,\phi}\approx  - \frac{{{\hbar ^2}}}{{2m}}\left( {\frac{{{\partial ^2}}}{{\partial r_{}^2}} + \frac{1}{{r_L^2}}\frac{{{\partial ^2}}}{{\partial \phi _{}^2}}} \right)$ due to the dominant $\frac{{{\partial ^2}}}{{\partial r_{}^2}}$ term.

In this paper, the ODT we study is composed of two red-detuned LG beams with a wavelength of 780 nm. The atom could be trapped at a maximum of laser intensity in each direction. As we have shown in Fig. 1(b), for a deep ODP, it is a good approximation to treat this ODT as a harmonic trap by means of second order perturbation. For this red-detuning induced ODT, We performed Taylor expansion at $r=r_L$ and $\phi=0$ respectively, and truncated it to the second order. The final form of ODT is:
\begin{equation}
{V_{\text{dip}}}\left( r \right) = \frac{{2\pi {c^2}\Gamma }}{{\omega _0^3}}\left( {\frac{1}{{{\Delta _1}}} + \frac{2}{{{\Delta _2}}}} \right)\sqrt L {I_{1,\max }}\left[ {1 - 2L{{\left( {\frac{r}{{{r_L}}} - 1} \right)}^2}} \right]\left( {1 - {L^2}{\phi ^2}} \right),\label{Eq 9}
\end{equation}
using the relation ${\left( {{r \mathord{\left/{\vphantom {r {{r_L}}}} \right.\kern-\nulldelimiterspace} {{r_L}}}} \right)^{2L}}{e^{ - L\left( {{r^2}/r_L^2 - 1} \right)}} \sim 1 - 2L{\left( {r/{r_L} - 1} \right)^2}$.

\begin{figure}[tbp]
\includegraphics[width=1\textwidth]{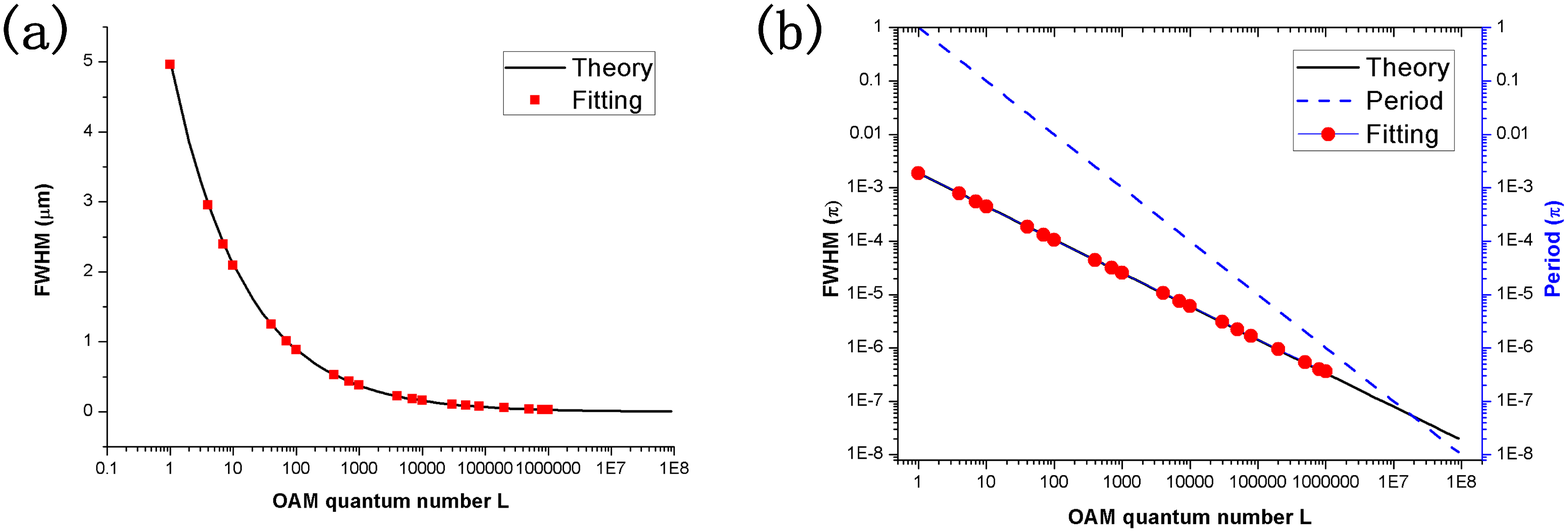}
\caption{\label{fig 1}(color online)The FWHM of ground states wave packets. (a) The FWHM in radial direction where $\phi=0$. (b) The FWHM of ODP in angular direction for $r=r_L$. In (a) and (b), The black solid lines and filled red circus are theoretical and numerical calculations, respectively. The blue dashed line in (b) is the width of single ODT in the $\phi$ direction. The data of blue dashed line are obtained from the approximate Hamiltonian.}
\end{figure}

Suppose the $^{87}$Rb atom is placed in the $z=0$ plane of the counter-propagating LG beams with wavelength $\lambda=780$ nm. With a fixed the toroid radius with $r_L=1$ mm. Which means that when we change the OAM quantum number $L$, Gaussian laser radius $w_0$ has to vary as $w_0=r_L\sqrt{2/L}$ synchronously. At the same time, in order to ensure the bound states of atoms in the ODT in the case of high $L$, we take the laser power as $P_0=10$ mW while the detuning form $D_2$ line $\Delta_2 \sim$21.483 GHz. For such near-resonance laser beams, the scattering effects of atomically stimulated absorption of photons and subsequent spontaneous reemission will not be negligible. One possible way to inhibit the spontaneous emission is to place the device in a cavity which all of normal mode are far off-resonance with respect to the atomic transition frequency \cite{Grimm00,Th98}.

According to the expression of ODT in Eq.(9) and our approximation, for a fixed radialus (azimuthal), the ODT in azimuthal (radial) direction could be regarded as a harmonic trap. The ground state in the harmonic trap is actually Gaussian shape. Thus, we calculate the full width at half maximum (FWHM) of this ground state Gaussian wave function, which can be written as:
\begin{equation}
\begin{aligned}
{\rm{FWH}}{{\rm{M}}_r}\left( L \right) = \frac{{2{r_L}}}{{\sqrt {\sqrt {\frac{{4r_L^2\pi {c^2}\Gamma }}{{\omega _0^3}}\left( {\frac{1}{{{\Delta _1}}} + \frac{2}{{{\Delta _2}}}} \right){I_{1,\max }}{L^{3/2}}m} /\hbar } }}\\
{\rm{FWH}}{{\rm{M}}_\phi }\left( L \right) = \frac{2}{{\sqrt {\sqrt {{L^{5/2}}\frac{{4r_L^2\pi {c^2}\Gamma }}{{\omega _0^3}}\left( {\frac{1}{{{\Delta _1}}} + \frac{2}{{{\Delta _2}}}} \right){I_{1,\max }}m} /\hbar } }},\label{Eq 10}
\end{aligned}
\end{equation}

In Fig. 2(a) and (b), we show the results of the theoretical model (filled red circus) and numerical calculations without harmonic approximation (black solid line) in the $r$ and $\phi$ direction. FWHM$_r(L)$ and FWHM$_{\phi}(L)$  vary as $L^{-3/8}$ and $L^{-5/8}$ with the increase of $L$. We only perform Gaussian fitting with wavefunction of ${\left| {\varphi \left( {r,0} \right)} \right\rangle _{\text{gs}}}$  and ${\left| {\varphi \left( {{r_L},\phi } \right)} \right\rangle _{\text{gs}}}$, where ${\left| {\varphi \left( {r,\phi } \right)} \right\rangle _{\text{gs}}}$ is the ground state of the atom in ODT.

\noindent
\textbf{IV. Sub-wavelength optical lattice with $\sim1$ nm period}

For the ring lattice we discussed in this article, there are $2L$ lobes in the azimuthal direction, the period of each lobe vary as $\pi/L$, we show this (the blue dashed line) in Fig. 2(b), as a comparison with FWHM of ground state wave packet in $\phi$ direction. When $L=3\times10^6$, the period $p_\phi$ of each fragment drops to 1 nm in the form of $p_\phi\sim\pi r_L/L$ where $r_L=1$ mm. At the same time, the waist of Gaussian beams is required as ~0.8 $\mu$m, which is well below the diffraction limit of the $780$ nm lasers, and is in principle feasible, since for a parallel beam of diameter $D$ passing through a convex lens with focal length $f$, it can be focused to a waist of $w=f\lambda /\pi D$.

In order to ensure the trapping efficiency of the ODT to atoms, the laser power $P_0$ is taken as $10$ mW. In Fig. 3(b), we demonstrated that the FWHM of ground state wave packet varies as $L\sim L^{-5/8}$ while the period of ODT in angular direction decrease as $\sim L^{-1}$. The $p$ drops faster than FWHM$_{\phi}$ as $L$ grows and there is also a cross between them, while FWHM$_{\phi}$ is much smaller than $p$ when $L=1$. And the purpose of the parameters we take is to make this cross point appear when $L>3\times 10^6$. As shown in Fig. 2(b), the period of the angular lattice decrease with $L^-1$, there is a limitation for this angular lattice of $L\sim 5\times 10^7$ for $P_0=10$ mW and $\Delta_2\sim21.483$GHz.

For $L=3\times 10^6$, we take the grid for coordinate $r$ in $r_L-80\sim r_L+140$ nm and for $\phi$ in $0\sim \pi/3\times 10^{-6}$, respectively. In order to simplify the calculation, we isolate one period of the optical dipole lattice, and discretize it on a $198\times99$ radial-angular mesh in the discrete variable representation (DVR) \cite{Colbert92}. We use the multi-configuration multi-layer time-dependent Hartree (ML-MCTDH) method \cite{Cao13,KCao13,KCao17} to calculate the spatial distribution of ground state $^{87}$Rb atom in the optical lattice (see Fig. 3). The calculated one-body density $\rho (r,\phi)$ proves the fact that the $^{87}$Rb atom can be tightly trapped in a small area of $ \sim$1 nm spacing in the angular direction.

\begin{figure}[tbp]
\includegraphics[width=1\textwidth]{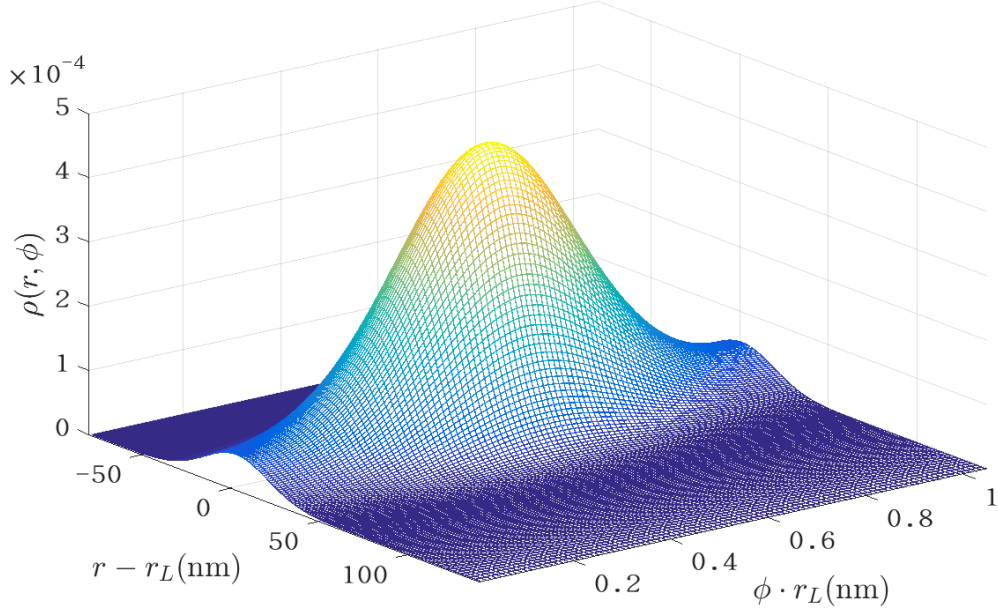}
\caption{\label{fig 3}(color online) The density  $\rho(r,\phi)$ of the bound ground state of $^{87}$Rb atom in the sub-wavelength optical dipole trap. Here, the radius $r_L=1$ nm and $\Delta r=1.11$ nm while $\Delta\phi=10^{-8}\pi$, correspondingly, for this fixed parameters $r_L$  and $\Delta \phi$ , the $\Delta r=10^{-2}$ nm  in tangential direction. The atom is mainly trapped in vicinity of the toroid equator with the range of $100$ nm  and $0.85$ nm  in radius and tangential direction.}
\end{figure}

\noindent
\textbf{Conclusion}

In this paper, we propose a scheme to create an optical lattice of $\sim1$ nm scale with two high order LG beams, with OAM quantum number $L=3\times10^6$. The radial distribution of dipole trap is monocyclic distribution with the width of $1$ mm. The $6 \times 10^6$ wells in angular direction, form an angular lattice with the period of $1$ nm. We calculate the one-body density of single $^{87}$Rb atom in a segment of the angular lattice and find that the atom can be trapped in a extremely narrow area. Since there is no fundamental limitation for the period to reach sub-nm scale, directly manipulating atoms to form molecules can be made possible, with further possibility of flexible 3D shaping of molecular structures based on 3D arrays of traps \cite{Barredo18}.

For a laser fields with optical modulation at nm scale, the non-dipole effect can be much more pronounced for the atoms interacting with proposed optical lattice, this will be adducted in the future work.\\

\noindent
\textbf{Acknowledgements}

We thank Li You, Jochen Kuepper and Henning Moritz for helpful discussions. This work was supported by the National Natural Science Foundation of China (Grants No. 11604107). Z.L. thanks Volkswagen Foundation for partial financial support through Peter Paul Ewald Fellowship.


\end{document}